\documentclass[aps,prb,amsmath,amssymb,twocolumn,superscriptaddress,floatfix,showpacs]{revtex4-1}
\usepackage{graphicx}
\usepackage{helvet}
\usepackage{hyperref}
\def\Tc{\ensuremath{T_\text{c}}}
\def\Hc2{\ensuremath{H_\text{c2}}}

\begin{document}
\title{Upper critical field of the noncentrosymmetric superconductor BiPd}

\author{Darren C.\ Peets}
\altaffiliation[Current address: ]{Advanced Materials Laboratory, Fudan University, Shanghai 200438, China}
\email{dpeets@fudan.edu.cn}
\affiliation{Max-Planck-Institut f\"ur Festk\"orperforschung,
 Stuttgart D-70569, Germany}
\affiliation{Center for Correlated Electron Systems,
 Institute for Basic Science, Seoul National University, 
 Seoul 08826, Korea}

\author{Ana Maldonado}
\affiliation{Max-Planck-Institut f\"ur Festk\"orperforschung,
 Stuttgart D-70569, Germany}
\affiliation{School of Physics and Astronomy, University of St.\ Andrews, North Haugh, St.\ Andrews, Fife KY16~9SS, United Kingdom}
\author{Mostafa Enayat}
\author{Zhixiang Sun}
\affiliation{Max-Planck-Institut f\"ur Festk\"orperforschung,
 Stuttgart D-70569, Germany}
\author{Peter Wahl}
\affiliation{Max-Planck-Institut f\"ur Festk\"orperforschung,
 Stuttgart D-70569, Germany}
\affiliation{School of Physics and Astronomy, University of St.\ Andrews, North Haugh, St.\ Andrews, Fife KY16~9SS, United Kingdom}

\author{Andreas P. Schnyder}
\affiliation{Max-Planck-Institut f\"ur Festk\"orperforschung,
 Stuttgart D-70569, Germany}

\date{\today}

\begin{abstract}
The superconducting parameters and upper critical field of the
noncentrosymmetric superconductor BiPd have proven contentious.  This
material is of particular interest because it is a singular example of
a $4f$-electron-free noncentrosymmetric superconductor of which
crystals may be grown and cleaved, enabling surface-sensitive
spectroscopies.  Here, using bulk probes augmented by tunnelling data
on defects, we establish that the lower of the previously reported
upper critical fields corresponds to the bulk transition.  The
material behaves as a nearly-weak-coupled BCS $s$-wave superconductor,
and we report its superconducting parameters as drawn from the bulk
upper critical field.  Possible reasons behind the order-of-magnitude
discrepancy in the reported upper critical fields are discussed.
\end{abstract}



\maketitle

\section{Introduction}

In noncentrosymmetric superconductors, the lack of spatial inversion
symmetry means that the parity of the superconducting wavefunction is
not a meaningful concept.  Singlet and triplet pairing states are a
consequence of parity, thus they are no longer eigenstates and can
mix.  With our usual simplifying assumptions for understanding
superconductivity no longer valid, a vast array of exotic physics
becomes possible\cite{Bauerbook,Fujimoto2007}.  Unfortunately, few
such materials are known, many do not superconduct under ambient
pressure, and single crystals have only been grown of a very few.  In
addition, significant spin-orbit splitting of the bands near the Fermi
level is a required prerequsite for all proposed novel behaviour.  As
a consequence of these challenges, most theoretical predictions remain
unrealized.

Known to be both superconducting\cite{Alekseevskii1952} and
noncentrosymmetric\cite{Kheiker1953} before BCS theory\cite{BCS1957},
$\alpha$-BiPd was probably the first noncentrosymmetric superconductor
identified as such.  That noncentrosymmetric superconductors were
particularly exotic is a far more recent discovery\cite{Bauer2004},
and this material is now attracting renewed attention, both as a
noncentrosymmetric
superconductor\cite{Joshi2011b,Okawa2012,Jiao2014,Jha2015,Yan2016} and
for its topologically nontrivial surface
states\cite{Wahl2015,Neupane2015}.  Several techniques have been
applied to establish whether the material hosts novel physics arising
from a mixed-parity condensate, including point-contact
spectroscopy\cite{Mondal2012}, NQR\cite{Matano2013}, and microwave
susceptibility\cite{Jiao2014}, revealing tantalizing hints of such
behaviour.  The gap symmetry and pairing mechanism in this material,
whether gap nodes occur, and the degree of parity mixing remain to be
established.  In fact, many of the material's most basic
superconducting parameters remain hotly contested.

Of particular concern, reports of the material's upper critical field
\Hc2, from which fundamental superconducting parameters such as the
coherence length are extracted, vary by more than an order of
magnitude.  Recent resistivity and {\it ac} susceptometry data suggest
an upper critical field \Hc2\ around
0.8\,T\cite{Joshi2011b,Mondal2012,Okawa2012}, while more
bulk-sensitive magnetization measurements and surface-sensitive
scanning tunneling spectroscopy indicate a far lower
value\cite{Joshi2014,Wahl2015}.  Where the values are similar, the
shape can disagree: {\it ac} susceptometry\cite{Mondal2012} leads to a
very different $H$--$T$ phase diagram than that extracted from the
resistive transition\cite{Joshi2011b}.  Here we present the bulk upper
critical field \Hc2\ as determined by magnetization, resistivity and
specific heat measurements, along with the superconducting parameters
that may be extracted based on this upper critical field.  Together
with tunnelling data, these results paint a consistent picture of a
singlet-dominated, fully-gapped pairing state, near weak-coupling BCS
expectations.  We discuss the likely reasons behind the discrepancy in
previous reports.

\section{Experimental}

Crystals were grown by a modified Bridgman-Stockbarger technique, as
described in greater detail elsewhere\cite{Peets2014}.  Chips of
bismuth (Aldrich, 99.999\%) and palladium metal (Degussa or Credit
Suisse, 99.95\%) in a stoichiometric ratio were sealed in an evacuated
quartz tube with a conical end, which was cooled slowly through the
material's congruent melting point
(600\,$^\circ$C\cite{BiPd,BiPd2006}) in a temperature gradient,
crystallizing the melt from one end of the tube at a growth rate of
1.5\,mm/h.  The ampoule was also cooled slowly through the transition
between $\alpha$-BiPd and $\beta$-BiPd\cite{Bhatt1979,Ionov1989} near
200\,$^\circ$C to maximize the domain size.  Reports thus far indicate
that the material always exists as $\alpha$-BiPd below this
transition, thus all measurements reported here were performed on the
$\alpha$ phase, which we refer to simply as BiPd.  The crystals were
twinned and in many cases internally cracked due to the
$\alpha$--$\beta$ phase transition, but otherwise single-domain.

Magnetization measurements were performed in Quantum Design MPMS-7 and
MPMS-XL magnetometers with the RSO option, and $\Hc2(T)$ was defined
as the point where the sample reached 10\% of its full, low-$H$,
low-$T$ magnetization in zero-field-cooled measurements.  Resistivity
was measured in a Quantum Design PPMS by a standard four-wire
technique on samples of approximate dimension
$2\times0.5\times0.5$\,mm$^3$, with a drive current of 5\,mA;
$\Hc2(T)$ was defined as the midpoint of the transition.  Note that
there may be systematic errors in the resistivity due to the
abovementioned cracking. To confirm the bulk, thermodynamic \Tc\ and
\Hc2, specific heat was measured at low temperatures in a Quantum
Design PPMS in zero field and for fields along the monoclinic $b$
axis.  In this case, $\Hc2(T)$ was found using an entropy-conserving
construction around the transition.  Point defects were investigated
using a home-built scanning tunnelling microscope (STM), operating in
cryogenic vacuum at temperatures below 30\,mK\cite{Singh2013}, on
samples that were cleaved {\it in situ} at low temperatures.

\section{Results}

\begin{figure}[htb] 
\includegraphics[width=\columnwidth]{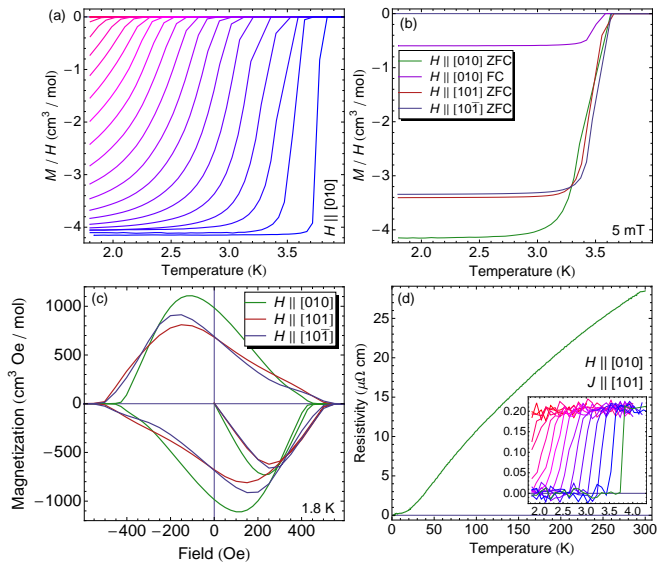}
\caption{\label{fig}(color online) Superconductivity in BiPd. (a)
  Zero-field-cooled magnetization curves in a field $H\parallel [010]$
  of 0.1\,mT, then every 2.5\,mT from 2.5 to 60.0\,mT (b)
  Magnetization curves in $H=5$\,mT for three field orientations; (c)
  $M$~--~$H$ loops at 1.8\,K for the three field orientations; and (d)
  zero-field resistivity --- the inset shows the effect of adding fields
  $H\parallel [010]$ in steps of 5\,mT.}
\end{figure}

Magnetization in a 5\,mT applied field is shown as a function of
temperature in Fig.~\ref{fig}(b) for several field orientations,
zero-field-cooled magnetization in a variety of fields $H\parallel
[010]$ is presented in Fig.~\ref{fig}(a), and $M(H)$ loops at 1.8\,K
are shown in Fig.~\ref{fig}(c).  The low-field transitions and $M(H)$
loops are essentially identical to those reported
elsewhere\cite{Okawa2012,Joshi2014,Jha2015}.  The material is
relatively isotropic, with the hysteresis loops exhibiting only minor
differences with field orientation.  We note that a recent report
claimed ferromagnetism in BiPd based on an $M$--$H$ just above
\Tc\cite{Jha2015} --- in our measurements, including in $M$--$H$ loops
for all three field orientations at the same measurement temperature
used in that report, no such hysteresis is observed and the
magnetization observed is several orders of magnitude weaker.  This
recent report is also inconsistent with all $M(T)$ data and with
previously published $M$--$H$ loops\cite{Joshi2014}, where such a
ferromagnetic signal would be a significant fraction of the
superconducting signal and clearly visible.  Resistivity is also
presented in Fig.~\ref{fig}(d), but absolute values may be unreliable
due to internal cracking resulting from the $\alpha$--$\beta$ phase
transition.  The residual resistivity ratio of 140 is comparable with
other work\cite{Joshi2011b,Jha2015}, as is the resistive \Tc\ of
3.8\,K.

\begin{figure*}[htb]
\includegraphics[width=\textwidth]{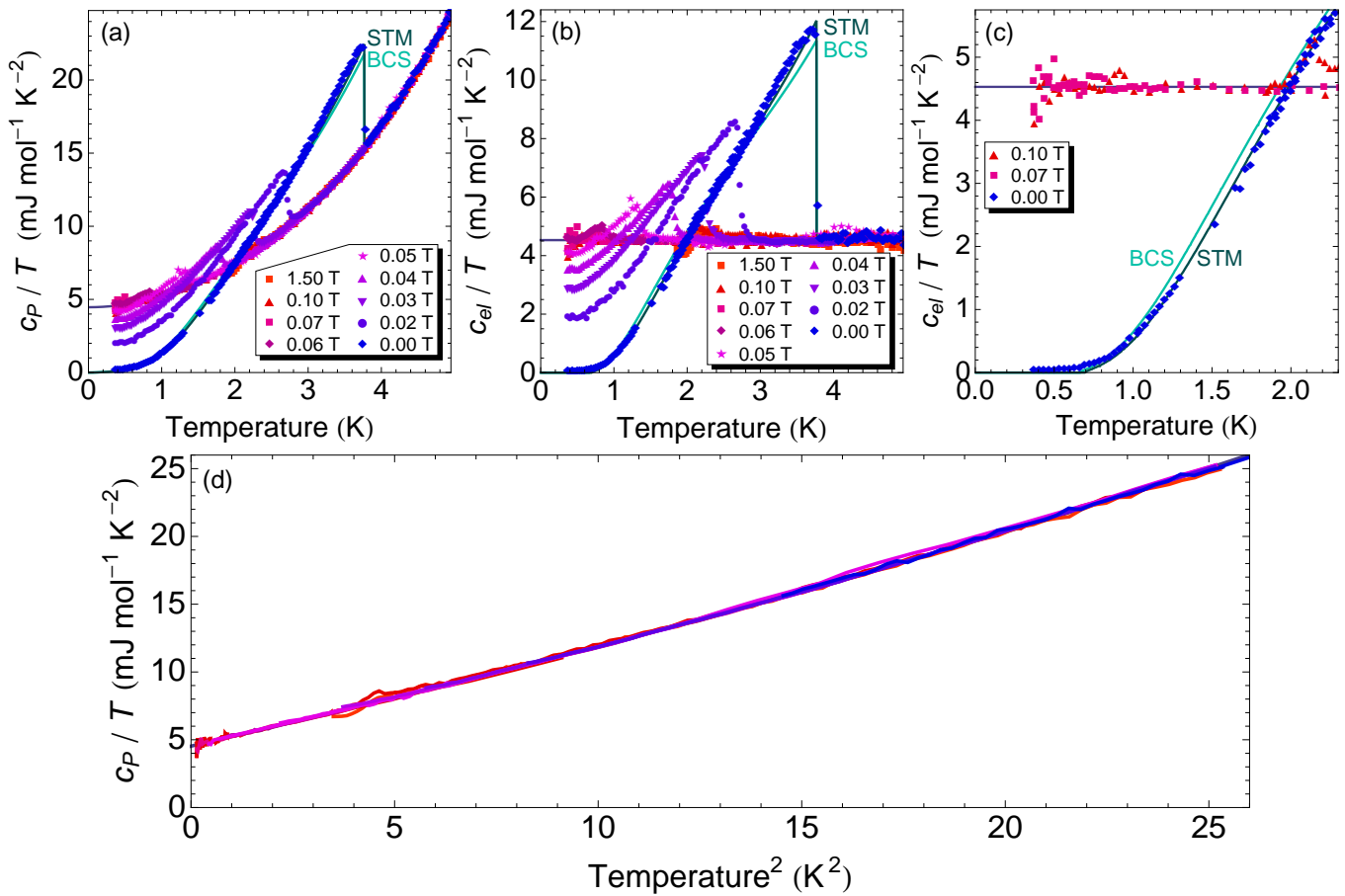}
\caption{\label{fig-cp}(color online) Specific heat of BiPd.  (a)
  $c_P(T)/T$ vs.\ $T$ for several fields, showing the bulk
  superconducting transition and its suppression by field.  The
  weak-coupling $s$-wave BCS form\cite{Muhlschlegel} and the BCS form
  recalculated using the STM gap\cite{Wahl2015} are included for
  comparison.  There is no transition visible above 0.4\,K in fields
  $H\geq 0.06$\,T.  (b) The electronic contribution $c_{el}(T)/T$
  vs.\ $T$. (c) The low-temperature region is fit better by the
  recalculated form based on the STM gap.  (d) All data in fields of
  70\,mT and above, and all data taken above the obvious
  (field-dependent) bulk transition, in all fields.  There is no
  evidence to suggest an additional phase transition.  The glitches
  near 4\,K$^2$ were not reproduced in later measurements.}
\end{figure*}

Specific heat was measured from 0.37 to 5\,K in fields up to 1.5\,T,
to determine the bulk superconducting transition and gain insight into
the structure of the gap function; results are presented in
Fig.~\ref{fig-cp}.  Describing the slight curvature visible in the
normal-state heat capacity in Fig.~\ref{fig-cp}(d) and isolating the
electronic contribution $c_{el}(T)$ required the addition of small
$T^5$ and $T^7$ corrections to the $T^3$ phonon term.  The zero-field
data are close to the weak-coupling BCS
expectation\cite{Muhlschlegel}, with a clear exponential onset
implying a full gap and suggesting a relatively isotropic gap
function.  As pointed out in our earlier paper\cite{Wahl2015},
recalculating the BCS form with the entropy-conserving \Tc\ and the
gap extracted from STM leads to a much better fit, without the need to
introduce anisotropy or additional gaps, but implying a small
deviation from weak coupling.  The jump height $\Delta
c_{el}/\gamma\Tc$ at \Tc\ is 1.50, slightly higher than the BCS
expectation of 1.43, and the \Tc\ of 3.77\,K agrees well with that
determined from the magnetization and resistivity measurements.
Magnetic field suppresses the transition, which is completely absent
above 0.4\,K in a field of only 0.07\,T.  It is worth noting at this
point that this agrees well with the data presented in Fig.~\ref{fig},
but implies an upper critical field more than an order of magnitude
lower than in the majority of recent reports.  Fig.~\ref{fig-cp}(d),
which depicts all specific heat data above the bulk transition
$\Hc2(T)$ for all fields, shows that there is no evidence for a second
phase transition which would correspond to the previously reported
transition.

\begin{figure}[htb]
\includegraphics[width=\columnwidth]{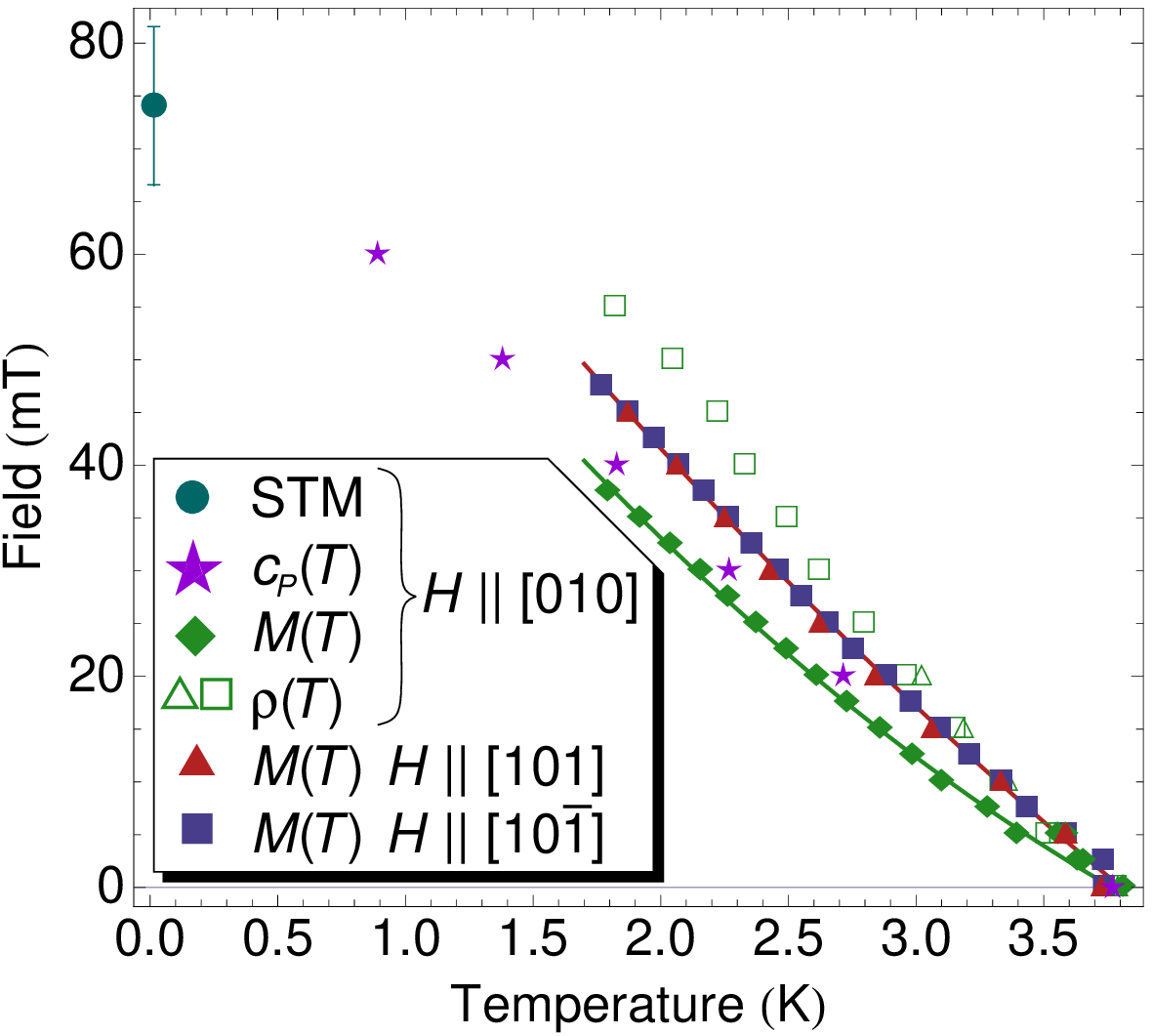}
\caption{\label{fig-HT}(color online) $H$ -- $T$ phase diagram
  constructed from the transitions in the resistivity, magnetization
  and specific heat.  A point corresponding to the closing of the gap
  as observed by STM\cite{Wahl2015} is included.}
\end{figure}

The transitions obtained from magnetization, resistivity and specific
heat measurements taken in a variety of applied magnetic fields may be
combined into an $H$~--~$T$ phase diagram, shown in Fig.~\ref{fig-HT}.
A low-temperature point from the closing of the gap detected by STM is
included\cite{Wahl2015}.  The slight upward curvature observed near
\Tc\ is common in multi-band systems, and indeed band structure
calculations indicate more than ten bands crossing the Fermi
level\cite{Wahl2015,Neupane2015}.  The data do not permit a reliable
extrapolation to \Hc2(0), but it is clearly the lower of the reported
values.  The discrepancy is discussed in greater detail below.

A variety of parameters characterizing BiPd and its superconductivity
may be extracted from the specific heat.  The Sommerfeld electronic
specific heat coefficient $\gamma$ is a modest 4.53\,mJ/mol\,K$^2$, and
the phonon $T^3$ term's prefactor $\beta$ is 0.710\,mJ/mol\,K$^4$,
corresponding to a Debye temperature of 176\,K.  The phonon
contribution's clear departure from $T^3$ behaviour even as low as
3\,K suggests that at least one phonon mode is rather low in energy.
The zero-field electronic specific heat $c_{el}$ may be integrated to
obtain the thermodynamic critical field $H_c$, using
\begin{equation}
\frac{\mu_0H_c^2}{2} = -\frac{\gamma\Tc^2}{2} + \int_{0}^{\Tc}c_{el}dT
\end{equation}
in appropriately chosen volume units.  For $H\parallel [010]$, the
resulting thermodynamic $H_c^b$ of 41\,mT, combined with the
$H_{c2}^b(0)$ of roughly 75\,mT, would indicate a
$\kappa_{GL}^{ac}\approx 1.3$.  This indicates weakly Type-II
superconductivity, near but not within the regime where the
intermediate mixed phase is possible.  The $H$--$T$ phase diagram is
remarkably isotropic for in-plane fields, so the $ac$-plane coherence
length $\xi^{ac}$ can be assumed to be isotropic to a good first
approximation, and the $\xi^{ac}(0)$ extracted from $H_{c2}^b(0)$ is
67\,nm.  If the penetration depth is similarly isotropic,
$\lambda^{ac}(0)=85$\,nm.  Because of the lower \Hc2, this penetration
depth is very far from the value reported previously and assumed in
calculating the microwave penetration depth\cite{Jiao2014}.

It is also possible to produce rough estimates of superconducting
parameters for in-plane field orientations.  On the assumption that
the factor of $\sim1.2$ anisotropy in Fig.~\ref{fig-HT} continues to
zero temperature, $H_{c2}^{ac}(0)$ would be 89\,mT, $\kappa_{GL}^{b}$
would take a slightly stronger Type-II value of 1.5, $\xi^b(0)$ would
be 56\,nm, and $\lambda^b(0)$ would be 103\,nm.  Confirmation of these
values will be necessary, either by extending measurements of $\Hc2^i$
to low temperature, or through independent measurements of the
coherence length and penetration depth, perhaps from muon spin
rotation.

\section{Discussion}

Upper critical fields on the order of 100\,mT have been reported based
on magnetization measurements\cite{Joshi2014,Jha2015,Wahl2015},
STM\cite{Wahl2015}, and now also specific heat; upper critical fields
closer to 1\,T have been inferred from
resistivity\cite{Joshi2011b,Okawa2012,Jha2015} and {\it ac}
susceptometry\cite{Mondal2012}, although resistivity at high drive
currents as reported here and in Ref.~\onlinecite{Wahl2015} appears to
support the lower value, and an intermediate resistive value has also
been reported\cite{Yan2016}.  Having concluded that the lower
\Hc2\ represents the bulk, thermodynamic transition, the question
arises as to the origin of the order of magnitude discrepancy.
Possibilities include a significant difference in samples between the
various groups, or some characteristic of the material that results in
the survival of weak, either surface or filamentary, superconductivity
above the bulk transition, which would short-circuit resistivity
measurements.

The first consideration to raise here is whether Bi vacancies could
alter the properties of the material.  It has been established that
BiPd can accommodate a considerable concentration of Bi vacancies,
although not without degradation of the
superconductivity\cite{Okawa2012}.  However, this degradation is not
rapid --- the removal of a staggering 22\%\ of all Bi atoms reduces
the residual resistivity ratio by a factor of 50 but \Tc\ only falls
by 35\%.  Perhaps more importantly, the resistive \Hc2\ appears to
track the reduction in \Tc, making an order-of-magnitude jump
unlikely.  There are also no obvious features in the calculated band
structure\cite{Wahl2015,Neupane2015}, such as a flat band, that would
suggest fundamental changes to the carriers for carrier concentrations
near the expected Fermi level.  The picture emerging, again, is that
of a relatively simple, BCS $s$-wave-like gap to first approximation.
In our crystals, EDX analysis suggested a slight Bi deficiency, but it
was within uncertainty of perfect stoichiometry.  A more precise EPMA
investigation was also consistent with ideal stoichiometry:
Bi$_{0.987(19)}$Pd$_{1.013(18)}$ using $2\sigma$ uncertainties.  Apart
from Ref.~\onlinecite{Okawa2012}, in which all samples were far from
stoichiometry, detailed information on the atomic ratios in the
crystals studied is not available.  However, the critical temperatures
vary little among the remaining works, suggesting only minor
deviations from ideal stoichiometry.

\begin{figure}[thb]
  \includegraphics[width=\columnwidth]{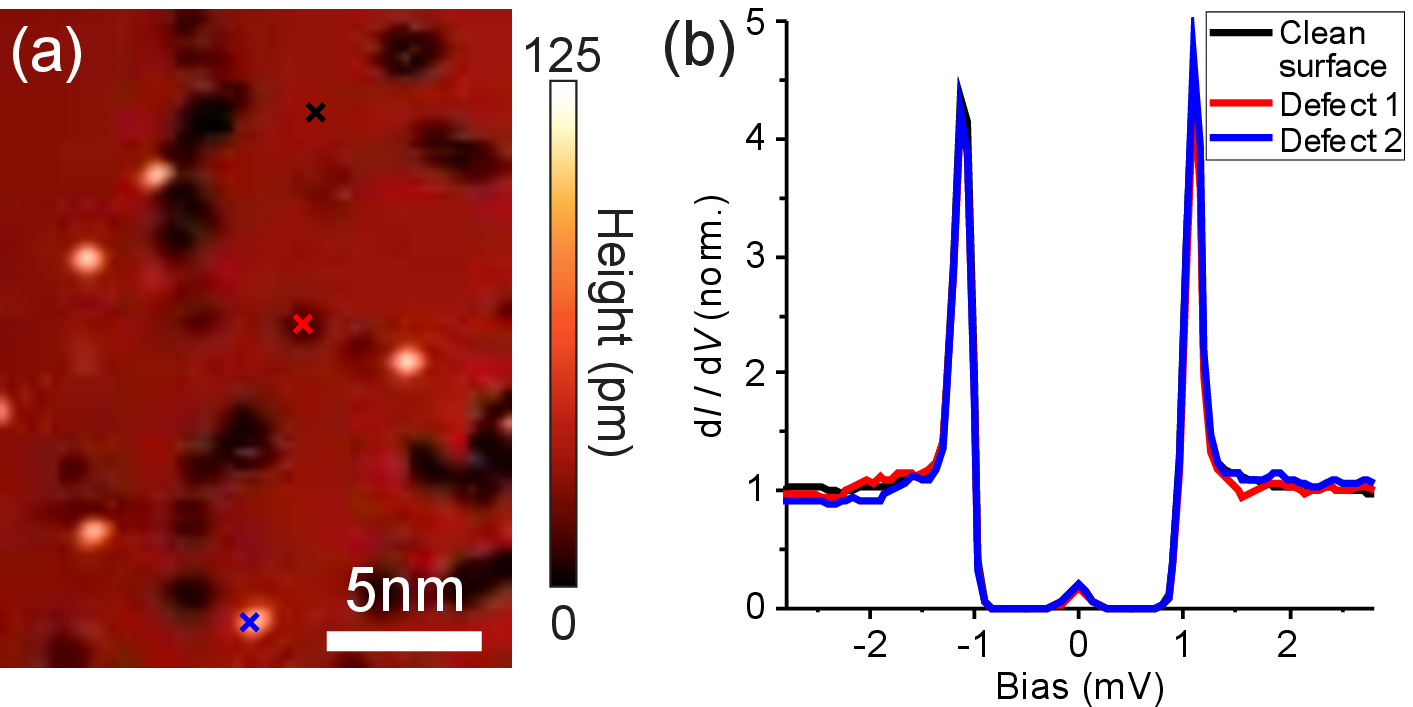}
  \caption{\label{fig:STM}(color online) Tunneling spectroscopy of
    individual defects. (a) Topographic image of defects in an
    atomically flat BiPd surface (settings: $V=50$\,mV,
    $I=0.1$\,nA). (b) Spectra taken on two different defects as well
    as the clean [010] surface, as marked in (a) with color-coded
    $\times$ signs (Defect 1 is dark). Spectra in panel (b) were
    obtained at 2\,K using a superconducting tip with a BiPd cluster
    at its apex, resulting in coherence peaks at $\pm2\Delta$. The
    small peak at zero bias occurs at the difference between the tip
    and sample gaps due to the finite measurement temperature.  An
    in-gap state would have produced an additional feature between the
    coherence peaks, which is not observed. The spectra on the defects
    show no obvious difference compared to the clean surface
    ($V=3.9$\,mV, $I=2$\,nA, $V_\mathrm{rms}=70$\,$\mu$V).}
\end{figure}

Having previously performed STM on these crystals\cite{Wahl2015}, we
also had access to topographic and spectroscopic information on the
observed point defects.  Based on our topographical scans ({\it
  e.g.}\ Fig.~\ref{fig:STM}a), point defects appear at approximately
the 0.5\%\ level per (surface) unit cell, or 0.25\% per site.

Figure \ref{fig:STM}(b) compares tunneling spectra collected on two
examples of point defects with spectra taken well away from them,
using a superconducting tip. In the absence of any magnetic atoms,
these defects can be expected to be non-magnetic scatterers.
Comparisons of point defects with defect-free regions were performed
at 15\,mK in an applied field of 35\,mT with a normal PtIr tip (not
shown) and at 2\,K in zero field with a superconducting BiPd tip.  In
both measurement modes, and on a range of distinct defects, spectra on
defects were indistinguishable from those measured far from defects,
and no in-gap states were observed.  The small peak at zero bias
voltage when using a superconducting tip arises due to the finite
temperature of the experiment --- tunnelling spectroscopy between two
superconductors at finite temperature yields a peak at a bias
corresponding to the difference of the two gaps, which increases in
height on increasing temperature. Since our tip and sample are both
BiPd, this peak appears at zero bias voltage.  That point defects are
indistinguishable from the bulk indicates that they aren't
pair-breaking and are not harmful to the superconductivity.  This is
consistent with a full, non-sign-changing pairing gap (minor
variations in phase would be possible).


Apart from point defects, the various samples could also differ in
their concentration of extended defects, most obviously either
inclusions or twin boundaries that occur at the $\alpha$--$\beta$
phase transition.  Any such defects would introduce scattering, but we
have already demonstrated that scattering does not have a strong
effect on the superconductivity.  $\beta$-Bi$_2$Pd has a higher
\Hc2\cite{Imai2012,Herrera2015}, but also a higher \Tc\ of 5.4\,K that
should be just as dominant in the zero-field resistivity as the higher
apparent \Hc2\ transition is under applied field and low drive
currents.  Suppressing its \Tc\ should also suppress its \Hc2, making
this an unlikely explanation.  To the authors' knowledge, the
low-temperature properties of Bi$_3$Pd$_5$ have not been reported.  A
wide variety of other Bi--Pd phases exist\cite{BiPd,BiPd2006}, but
they should not be able to form as inclusions in a BiPd sample.

The crystals here were cooled slowly through the $\alpha$--$\beta$
structural transition, while other groups typically cooled rapidly
through this temperature range, which should lead to significant
differences in the concentration of twin boundaries and the spatial
distribution of strain.  The twin domain size could be smaller than
the coherence length in some samples, perhaps making the material
effectively centrosymmetric or leading to a reduced, effective
coherence length.  We cannot test this in other groups' crystals, but
in STM work on our samples, the considerable difficulty in locating a
twin boundary would strongly suggest a domain size well in excess of
the zero-temperature coherence length, at least within the $ac$-plane.
However, since the coherence length is strongly temperature-dependent,
crystals with twin domains large compared to $\xi$ at low temperature
would be in the opposite regime closer to \Tc.  If any crossover
existed between two regimes that differed by an order of magnitude in
critical field, we would expect this crossover to be clearly visible.
It is not.

The remaining possibility is that the bulk transition is masked from
some measurement techniques --- regions of many samples are clearly
still superconducting well above the bulk \Hc2.  A surface critical
field ($H_{c3}$) can explain a higher apparent transition in
techniques sensitive to the sides of the sample where the applied
field is parallel to the surface.  In Ginzburg-Landau theory for a
single-component order parameter, $H_{c3} = 1.695\Hc2$, which would
remain nearly an order of magnitude short of explaining the
discrepancy in upper critical fields.  In noncentrosymmetric
superconductors, however, the mixed-parity condensate is not described
by a single-component order parameter, and the constraints on $H_{c3}$
would need to be rederived.  Surface critical fields exceeding
1.695\Hc2\ have been invoked in discussing discrepancies between
resistivity and more bulk-sensitive probes in noncentrosymmetric
LaRhSi$_3$\cite{Kimura2016}, LaNiC$_2$\cite{Hirose2012},
LaIrSi$_3$\cite{Anand2014}, and LaPdSi$_3$\cite{Smidman2014}.
However, a surface critical field cannot explain the thermal transport
results on BiPd\cite{Yan2016}, unless the `surface' comprised
$\sim$30\%\ of their cubic-millimeter-scale crystal.



The higher-field transition could also be filamentary, but the
superconducting filaments must be sufficiently interconnected to
permit lossless electrical transport, respond to $ac$ susceptometry,
and prevent $\sim$30\% of the carriers from participating in heat
transport just above the bulk \Hc2\cite{Yan2016}.  The obvious network
of extended defects throughout each sample is twin boundaries.  Being
two-dimensional, these would support noticeable supercurrents, would
be relatively well-connected throughout the sample, and would have
enhanced apparent critical fields for fields oriented within the plane
of the twin boundary, due to being in the thin limit.  The loss of
carriers in thermal transport could be explained by superconducting
domain boundaries walling off areas of the sample or otherwise
blockading the flow of heat --- if 30\%\ of the material were still
superconducting, there should be clear signatures in the specific
heat.  Our \Hc2\ value would place the BiPd thermal transport data
among the $s$-wave superconductors.

\section{Conclusion}

The data presented here paint a comprehensive picture of BiPd as a
single-gap, nodeless, dominantly-$s$-wave BCS superconductor, albeit
with slight deviations from weak coupling.  The triplet component is
apparently not strong enough to lead to significant gap anisotropy or
nodes, as these would be seen in the specific heat.  The upper
critical field exhibits upward curvature suggestive of multi-band
superconductivity, which is unsurprising given that band structure
calculations show approximately 13 bands crossing the Fermi
level\cite{Wahl2015,Neupane2015}.  Previously reported values of
\Hc2\ disagree by more than an order of magnitude, and our results
indicate that the lower values reflect the intrinsic behavior;
superconducting parameters based on this are summarized above.  Since
parameters based on the higher \Hc2\ have been assumed in ensuing work,
it would be desirable to recalculate some quantities, notably the
microwave penetration depth\cite{Jiao2014}.  The higher \Hc2\ values
are most likely attributable to filamentary superconductivity
occurring in thin regions along twin boundaries, although this remains
to be demonstrated.  As for {\sl why} this would happen, strain is one
candidate, but we note that isostatic pressure suppresses the
superconductivity\cite{Jha2015}.  In Sr$_2$RuO$_4$, widely interpreted
to be a chiral $p$-wave superconductor, the superconducting onset
temperature can double at boundaries with Ru metal
inclusions\cite{Maeno1998}, and there is some evidence that this
boundary may nucleate one component of the multi-component order
parameter\cite{Nakamura2011}.  The parity mixing in noncentrosymmetric
superconductors implies that their condensates are effectively
multi-component, so twin boundaries may exhibit analogous physics in
BiPd.

\section*{Acknowledgements}
The authors thank C. Stahl and E. Goering of Abt. Sch\"utz, MPI-IS,
for help with magnetization measurements, X.\ P.\ Shen for the EPMA
measurements, and H.\ Jo for help measuring low-temperature specific
heat.  The authors are also grateful for the assistance of the MPI-FKF
Crystal Growth and Chemical Service Groups.  Work at IBS-CCES and SNU
was supported by the Institute for Basic Science (IBS) in Korea
(IBS-R009-G1), and work at St.\ Andrews was supported by the EPSRC
through grant EP/I031014/1.

Data associated with this work are available in
Ref.~\onlinecite{PRBrawdata}.

\bibliography{BiPd_arXiv}
\end{document}